\documentstyle[prb,tighten,eqsecnum,aps,twocolumn]{revtex}
\begin{document}
\title{The two dimensional antiferromagnetic Heisenberg model in the
presence of an external field}
\author{M.S.~Yang 
                and  K.H.~M\"utter}
\address{Physics Department, University of Wuppertal, D-42097 Wuppertal, Germany} 

\date{\today}
\maketitle
%
%
%
%
\begin{abstract} \\
We present numerical results on the zero temperature magnetization curve
and the static structure factors of the two dimensional antiferromagnetic 
Heisenberg model in the presence of an external field. The impact of frustration
is also studied.
\end{abstract}
\draft
%
\section{Introduction}
%
%
The discovery of high-T$_{c}$ superconductors \cite{Bednorz86} has renewed the interest
in the two dimensional antiferromagnetic spin-$\frac{1}{2}$ Heisenberg
Hamiltonian :
\begin{equation}
{
H = \sum_{<x,y>} \vec{S}(x) \vec{S}(y) + \alpha \sum_{[x,y]} \vec{S}(x) \vec{S}(y).
}
\end{equation} 
The ground state of (1.1) with nearest neighbors couplings $<\!\!x,y\!\!>$
is considered to yield a good description of the antiferromagnetic 
properties in the undoped copper oxide planes. Doping destroys the
antiferromagnetic order and this effect has been studied \cite{Doniach88,Schulz92} 
intensively by adding to the
nearest neighbor Hamiltonian a second term with next to nearest neighbor
couplings $[x,y]$ in diagonal directions of strength $\alpha$=$J_{2}/J_{1}$.
The behavior of the static structure factors $S_i(\vec{p}=(p_{1},p_{2}),\alpha,N)=
4 \sum \exp(i \vec{p} \vec{x})\!\!<\!\!S_i(0) S_i(\vec{x})\!\!>$ -
which are just the Fourier transform of the spin-spin correlators 
$<\!S_i(0) S_i(\vec{x})\!>$, i=1,2,3 - changes with $\alpha$ : \\
\\
1.) For 0 $<$ $\alpha$ $<$ 0.3 one finds a strong divergence
\begin{equation}
{  
S_i(\vec{p}=(p_{1},p_{2}),\alpha,N \to \infty) =  4 N m^{+^{2}}(\alpha)
}
\end{equation} 
at the momentum $\vec{p}=(\pi,\pi)$, indicating antiferromagnetic 
order with a non vanishing staggered magnetization $m^{+}(\alpha=0) \simeq$0.308.
\cite{Schulz92} \\
\\
2.) For $\alpha$ $>$ 0.7 such a divergence is found at
$\vec{p}=(0,\pi)$ and $\vec{p}=(\pi,0)$, which is a signature for collinear 
antiferromagnetic order. \\
\\
It is unclear, what happens in between 0.3 $<$ $\alpha$ $<$ 0.7. One
can imagine, that the singularity is less divergent and starts to move along some
trajectory from $\vec{p}=(\pi,\pi)$ to $\vec{p}=(\pi,0)$ and
$\vec{p}=(0,\pi)$.
The reconstruction of this trajectory would demand a very precise
determination of the static structure factor $S_i(\vec{p},\alpha,N)$ with a high
resolution in the momentum plane. \\
Moving singularities in static
structure factors were observed and analyzed for the one dimensional
antiferromagnetic Heisenberg model in the presence of an external field $B$. 
\cite{Johnson86,Karbach95} The relevant ground states, 
which enter in the computation of the static
structure factors have total spin $S_{3}$=$S$=$M(B) N$, where $M$=$M(B)$ is
given by the magnetization curve. 
The following phenomena have been found in the one dimensional 
antiferromagnetic Heisenberg model at fixed magnetization $M$ : \\
\\
1) the longitudinal structure factor is almost constant for 
$p_{3}(M)=\pi (1-2M) < p \le \pi $ : 
\begin{equation}
S_{3}(p,M) \simeq 2(1-2M)
\end{equation}
and develops a cusp type singularity \cite{Karbach95} at $p=p_{3}(M)$.
The critical exponent $\eta_{3}(M)$ at this singularity changes with 
the external field \cite{Fledder96} - i.e. with M. \\
\\
2) The transverse structure factor is almost constant for $0 \le p < 
p_{1}(M)=2 \pi M $ : 
\begin{equation}
S_{1}(p,M) \simeq 2 M
\end{equation} 
and develops a ''break'' type singularity
at the momentum $p_{1}(M)$. Moreover, it diverges at
$p=\pi$ with a field-dependent critical exponent $\eta_{1}(M)$. \cite{Fledder96}\\
\\ 
The singularities in the static structure factors are accompanied with 
zero frequency excitations (''soft-modes'') in the dynamical structure factors. 
\cite{Fledder96,Mueller81}
The effect of frustration on the position 
$p_{3}(M)$, $p_{1}(M)$ and the type $\eta_{3}(M)$, $\eta_{1}(M)$ of
the singularities has been studied as well. \cite{Schmidt96} 
Frustration does not change the position but drastically changes the type of the 
singularities. \\
\\
It is the purpose of this paper to investigate the singularities in
the static structure factors of the two dimensional spin-$\frac{1}{2}$ Heisenberg model in 
the presence of an external field. \\
Our analysis is based on a numerical
computation of the ground states at fixed magnetization $M$=$S/N$ on
the following square lattices : 
\begin{equation}
N = 4\times 4 , \;\;\;\; N = 6\times 6
\end{equation}
with periodic boundary conditions and
\begin{equation}
N = 3\times 3 \; \underline{+} 1 , \;\;\; N = 5\times 5 \; \underline{+} 1 , \;\;\; 
N = 7\times7 \; \underline{+} 1
\end{equation}
with helical boundary conditions. \cite{Haan92}
In the latter case the Hamiltonian can be considered to be one dimensional
\begin{eqnarray}
H &=& \sum_{x=1}^N \vec{S}(x) (\vec{S}(x+1)+\vec{S}(x+k))  \nonumber \\
  &+& \alpha \left( \sum_{x=1}^N \vec{S}(x) (\vec{S}(x+k-1)+\vec{S}(x+k+1))\right) 
\end{eqnarray}
with four types of couplings. Square lattices are realized for : \\
\begin{equation} { N=k^2 + 1 , \;\; k=3,5,7 } \end{equation}
\begin{equation} { N=k^2 - 1 , \;\; k=3,5,7 } \end{equation}
as can be seen for $k$=$5$, $N$=$26$ in Fig. 1.\\
\\
The helical boundary conditions yield a quantization of the two dimensional
momenta $\vec{p}=(p_{1},p_{2})$ : 
\begin{equation} 
{
p_{1}=\frac{2 \pi}{N} n , \;\;\; p_{2}=\frac{2 \pi}{N} k n , \;\;\; 
n=0, ... ,\frac{N}{2} 
}
\end{equation}
which differs from the usual quantization due to 
periodic boundary conditions. \\
\\
The outline of the paper is as follows. In section 2 we present the 
magnetization curves $M(\alpha,B)$ at zero temperature, as they depend on the
external field $B$ and the frustration parameter $\alpha$. The field-dependence of the
static structure factors at momentum $\vec{p}=(\pi,\pi)$ is discussed in
section 3. The momentum dependence of the static structure factors turns out 
to be smooth for the unfrustrated model, as is demonstrated in section 4.
Therefore, we did not find any signature for field-dependent soft-modes 
in this case ($\alpha=0.0$).
The situation changes if we frustrate the system sufficiently ($\alpha=0.5$). 
There we observe cusp-type singularities in the field-dependence of the static
structure factors at certain momenta (section 5).
%
\section{The magnetization curve}
%
Let us start with the ground state energies per site  
$\epsilon(\alpha,M$=$\frac{S}{N},N)$ at fixed magnetization $M$=$\frac{S}{N}$ 
and frustration $\alpha$. The numerical results for the 
systems (1.5) and (1.6) scale with $M$ except for the vicinity of 
$M$=$0$ and $M$=$\frac{1}{2}$. 
Near saturation $M \to \frac{1}{2}$,  
the ground state energy per site 
\begin{equation}
{
\epsilon(\alpha,M=\frac{1}{2},N)=\frac{\alpha+1}{2}
}
\end{equation}
and its first derivative 
\begin{equation} 
{\frac{\epsilon(\alpha,M=\frac{1}{2}-\frac{1}{N},N)-
 \epsilon(\alpha,M=\frac{1}{2},N)}{M-\frac{1}{2}}=4,
}
\end{equation} with
$\alpha \leq \frac{1}{2}$ are known for all system sizes $N$. \\
Finite size effects appear in the difference ratios :
\\
\begin{eqnarray}
(M-\frac{1}{2})^{-1} (\epsilon(\alpha,M,N)-\epsilon(\alpha,M=\frac{1}{2},N)) \nonumber \\
= D(\alpha,Z) \nonumber \\
= 4 - \frac{1}{2} D_{1}(\alpha) Z - \frac{1}{2} D_{2}(\alpha) Z^2 - 
\frac{1}{2} D_{3}(\alpha) Z^3 
\end{eqnarray}
\\
They can be absorbed in the definition of an ''optimized'' scaling variable :
\begin{equation}
{
Z=(1-2M)-\frac{c(M)}{N}
}
\end{equation}
with 
\begin{equation}
c(M)=2 (\frac{1/2-1/N}{M})^{\kappa}. 
\end{equation} 
The choice for $c(M)$ guarantees, that the expansion (2.3) satisfies the ''initial''
condition (2.2) for $M$=$\frac{1}{2}-\frac{1}{N}$ i.e. Z=0 independent of the
exponent $\kappa$. For $\kappa \simeq 5$, we achieve the best scaling of the
numerical data as can be seen from Fig. 2.
A fit of the numerical data to a Taylor expansion in $Z$ yields the coefficients 
$D_{n}(\alpha), \;\; n=1,2,3$ listed in table 1. We have repeated the fit 
with one additional term in the Taylor expansion (2.3). Such a fit yields only
slight modifications of the first two coefficient.\\
\\
It is remarkable to note, that the first coefficient $D_{1}(\alpha)$ 
decreases with $\alpha$ and vanishes at some value $\alpha$=$\alpha^{*}$
where 0.49 $<$ $\alpha^{*}$ $<$ 0.5. This has important consequence for
the magnetization curve $M$=$M(B,\alpha)$ which is obtained from 
$\epsilon(\alpha,M,N=\infty)$ by differentiation : 

\begin{equation}
{
\frac{\partial{\epsilon}}{\partial{M}}=B(M).
}
\end{equation}
For $0 < \alpha < \alpha^{*}$, where  
$D_{1}(\alpha) > 0$, we derive from (2.3) and (2.6) a linear behavior 
near saturation $M \to \frac{1}{2}$ 
\begin{equation} 
{ 
M=\frac{1}{2} - \frac{4-B}{D_{1}(\alpha)} 
} 
\end{equation} 
For $\alpha=\alpha^{*}$ - where $D_{1}(\alpha^{*})=0$ - 
however, we find a square root behavior :
\begin{equation} 
{ 
M=\frac{1}{2} - \sqrt{\frac{4-B}{2 D_{2}(\alpha^{*})}} 
} 
\end{equation} 
A similar phenomenon has been observed \cite{Schmidt96} 
in the one dimensional antiferromagnetic 
Heisenberg model with frustration $\alpha$. There it turns out, 
that the magnetization curve has a square root behavior of the type (2.8)
in the spin fluid phase 
$0 \leq \alpha < \alpha_{0}$ with $\alpha_{0} \simeq 0.241 $ and
a quartic root behavior in the vicinity of the transition point $\alpha$,
namely at $\alpha$=0.25.\\
\\
Let us go back to the discussion of the 
magnetization curves in the two dimensional model.
The coefficients $D_{n}(\alpha), \; n=1,2,3$ in the polynomial (2.3) were
fitted by the data points in the regime $Z < \frac{1}{3} \;\; 
(\frac{1}{3}<M<\frac{1}{2})$.
The extrapolation of (2.3) to larger $Z$-values (smaller $M$ values) is close
to and away from the numerical data for $\alpha$=0.0 and $\alpha$=0.5, 
respectively. In other words : $D(\alpha=0.0,Z)$ seems to be a
smooth function in $Z$ in contrast to $D(\alpha=0.5,Z)$ which changes its slope
substantially somewhere between $Z$=0.3 and $Z$=0.4. In order to determine
the exact position $Z$, where the slope may be discontinuous, we would need
numerical results on larger systems, which allow for a better resolution 
of the $Z$ dependence of (2.3). A discontinuity in the slope of $D(\alpha=0.5,Z)$
at $Z=Z_{o}$ (i.e. $M=M_{o}$) means, that the magnetization curve (2.6) has a
plateau at height $M=M_{o}$.\\ 
The existence of a plateau in the magnetization curve for $\alpha >$ 0.5 
has been predicted by S. Gluzman \cite{Gluzman94}, 
who studied the equivalent Bose gas problem and found a gap.
In Figs. 3a,b we present the magnetization curves for $\alpha=0.0$ and
$\alpha=0.5$ as they follow from an analysis of finite system results with the
method of Bonner and Fisher. \cite{Bonner64} Finite size effects appear to be small for
$\alpha=0.0$, as can be seen from a comparison of the numerical data for
the small system sizes ($N=24,26$) with those for larger systems 
($N=36,48,50$) shown in the insets of Fig. 3a. \\
For $\alpha=0.5$, however, we observe a sensitivity of the numerical data
to the size of the system and to the boundary condition. The results for
$N=24,26$ obtained with helical boundary conditions have a shoulder around
$M=\frac{1}{3}$, which might indicate the emergence of a ''plateau''. 
Unfortunately, we do not reach this shoulder
with our largest systems $N=48,50$ shown in the upper insets of
Fig. 3b. The lower inset of Fig. 3b contains the results for $N=36$ 
with periodic boundary conditions. Here, we observe a shoulder around $M$=$0.42$
which seems to be absent in the upper insets for the systems with $N$=$48,50$.
The variation of the magnetization curve with $\alpha$ on the $4\times 4$ system
has been computed by Lozovik and Notych.\cite{Lozovik93} These authors claim to see indications for ''plateaus'' at various $M$-values for $\alpha \simeq$ 0.538.
%
\section{The field-dependence of the static
structure factors at $\vec{p}$=$(\pi,\pi)$}
%
In the one dimensional spin-$\frac{1}{2}$ antiferromagnetic Heisenberg model,
the field-dependence of the longitudinal and transverse structure factors at
$p$=$\pi$ looks very different \cite{Karbach95} : Finite size effects are
small in the longitudinal case, the data points nicely scale with
$M$ (for $M > 0$) and develop a logarithmic singularity for $M \to 0$. Finite
size effects are very strong in the transverse case, the data points
do not scale at all with M and indicate that the singularity persists at
$p$=$\pi$ if the external field is switched on. The same 
behavior can be seen in the two dimensional case. In Fig. 4 we present
the longitudinal structure factors $S_{3}(\vec{p}=(\pi,\pi),M,\alpha,N)$ with 
$\alpha=0.0$ and $\alpha=0.5$. All the data points follow one 
unique curve.
Finite size effects turn out to be very small for larger values of $M$. 
For decreasing values of $M$ however, we find increasing finite-size
effects. This is a signature for the emergence of a singularity in the limit
$M \to 0$. Note also, that the longitudinal structure factor decreases 
with the frustration parameter $\alpha$. \newline
The transverse structure factors $S_{1}(\vec{p}=(\pi,\pi),M,\alpha,N)$
are shown in Fig. 5a and Fig. 5b for $\alpha$=$0.0$ and $\alpha$=$0.5$, respectively. 
The strong finite-size dependence for $\alpha$=$0.0$ indicates, that a 
weak magnetic field strengthens the
singularity at $\vec{p}$=$(\pi,\pi)$ and thereby the antiferromagnetic order in
the transverse structure factor. Note also the drastic change when we
pass from the unfrustrated case (Fig. 5a) to the frustrated
case (Fig. 5b).
In the latter case the finite-size dependence is less pronounced. 
Moreover, there appears a dip at $M$=$\frac{1}{3}$ in the data for
$N=24, 26$ and at $M$=0.42 for $N=36$ (inset Fig. 5b). In both
cases, this is just the position, where we found a shoulder in the
corresponding magnetization curves. \\ The dip at $M$=$\frac{1}{3}$ has a
good chance to survive in the thermodynamical limit; it is at least 
consistent with the numerical results on $N=24,26,36,48,50$. The dip
at $M$=0.42 in the data points seems to be a mere finite-size and/or
boundary value effect. 

%
\section{The momentum dependence of the static structure factor 
in the unfrustrated model ($\alpha=0$)}
%
It was pointed out in the introduction that the static structure factors
of the \it{one dimensional} \normalsize spin-$\frac{1}{2}$ antiferromagnetic Heisenberg model
develop singularities at field-dependent momenta ($p_{1}(M)$=$2 \pi M$,
$p_{3}(M)$=$\pi(1-2M)$). We therefore address here the question, whether such 
singularities can be found in the momentum dependence of the static
structure factors for the two dimensional model as well. It will turn out, that
the appearance of these singularities crucially depends on the
frustration parameter $\alpha$. In this section we first treat the 
unfrustrated case $\alpha$=$0.0$.\\ 
\\
The magnetization $M$=$\frac{5}{12}$ is realized on three systems (1.5) (1.6) with
N=24, 36, 48. \\ The numerical values for the longitudinal
structure factor $S_{3}(\vec{p}=(p_{1},p_{2})$,$M$=$\frac{5}{12},
\alpha=0,N)$, $N$=$24,36,48$ in the first 
Brillouin zone can be read off Fig. 6. These values appear to be constant along 
lines of constant :
\begin{equation}
{
x=\cos{p_{1}}+\cos{p_{2}} 
}
\end{equation}
represented by the dashed curves in Fig. 6. The behavior can be seen 
directly along the line $p_{1}+p_{2}$=$\pi$, $x=0$. Therefore, we
expect the longitudinal structure factor to depend
on the variable $x$ only.
In Fig. 7 we have plotted the longitudinal structure factor at $M$=$\frac{1}{3}$
$(N=24,36)$, $M$=$\frac{3}{8}$ $(N=24,48)$ and  
$M$=$\frac{5}{12}$ $(N=24,36,48)$ versus $x$. We observe a constant 
behavior for $-2<x<x_{3}(M)$, and afterwards a
decrease for $x_{3}(M)<x<2$. $x_{3}(M)$ increases with $M$.\\
\\
The corresponding $x$-dependence of the transverse structure factor is shown
in Fig. 8. For $x \to -2 $ $(\vec{p} \to (\pi,\pi))$ we see here the
emergence of a singularity in accord with the strong finite-size dependence of the
transverse structure factor $S_{1}(\vec{p}=(\pi,\pi),M,\alpha=0,N) -2M$
discussed in section 3. Moreover, this quantity is almost zero for 
$x > x_{1}(M)$, where $x_{1}(M)$ decreases with $M$.
At $M$=$0$ the longitudinal and transverse structure factor coincide. Their
dependence on the variable (4.1) is shown in Fig. 9 for $N=24,26,36$.\\
\\
In the one dimensional case the longitudinal and transverse structure factors turned
out to be constant for $p>p_{3}(M)$=$\pi(1-2M)$ and $p<p_{1}(M)$=$\pi 2M$,
respectively. However, there one finds cusp-type singularities at the 
''soft mode'' momenta $p$=$p_{3}(M)$ and $p$=$p_{1}(M)$. The smooth
momentum dependence of the static structure factors for the
\it{unfrustrated  two dimensional} \normalsize Heisenberg model seems to exclude
the existence of field-dependent soft-modes.   
%
\section{hunting for soft-modes}
%
In one and two dimensions, the antiferromagnetic Heisenberg models with 
nearest neighbor couplings differ substantially under various aspects : \\
\\
1. Fluctuations are stronger in one than in two dimensions i.e. the
   antiferromagnetic order increases with the dimension. \\ 
\\
2. There are field-dependent soft-modes in one dimension; they seem to be absent
   in two dimensions. \\
\\
3. Near saturation ($B \to 4, M \to \frac{1}{2}$) the magnetization curves
show a linear behavior in two dimensions but a square root behavior in
one dimension. As was pointed out in section 2, this behavior changes, if
the frustration parameter exceeds a critical value $\alpha^{*}$.\\
\\
If these features are correlated, we might hope to find soft-modes in the
two dimensional antiferromagnetic Heisenberg model as well, provided we 
weaken the antiferromagnetic order by frustrating the system. 
In order to test this hypothesis, we need a reliable
criteria for the existence or nonexistence of soft-modes. \\
\\
In the one dimensional case \cite{Karbach95}, 
we looked for singularities (breaks, cusps etc) in
the momentum distribution of the structure factors at fixed magnetization $M$.
As is demonstrated in Figs. 10a,b for the one dimensional case, pronounced structures
are produced as well by the soft-modes in the $M$-dependence. The position
of these singularities $M_{1}(p)=\frac{p}{2 \pi}$ and $M_{3}(p)=\frac{\pi-p}{2 \pi}$
define the soft-mode trajectories, which travel from $p=0$ (ferromagnetic order)
to $p=\pi$ (antiferromagnetic order) and vice versa. Keeping fixed the momentum,
we have to meet the soft-mode at an appropriate value of the magnetization.
In this respect it is much easier to find the soft-modes in the one dimensional case.\\
In case of the frustrated two dimensional Heisenberg model, the points 
$\vec{p}=(\pi,\pi)$  and $\vec{p}=(\pi,0)$, $\vec{p}=(0,\pi)$ play a special
role. Singularities in the structure factors at these points indicate 
antiferromagnetic and collinear antiferromagnetic order, respectively. Therefore,
it is rather plausible to assume, that the soft-mode trajectories connect these
points. \\
On the system with $N = 5\times 5 -1= 24$ sites, the momentum closest to 
$\vec{p}=(\pi,\pi)$ and $\vec{p}=(0,\pi)$ are $\vec{p}=\pi 
(\frac{2}{3},\frac{2}{3})$ and $\vec{p}=\pi (\frac{1}{6},\frac{5}{6})$,
respectively. Here, we have the best chance to observe soft-mode effects in the
$M$-distributions of the static structure factors. The sensitivity of the
$M$-distributions to a change of the frustration parameter can be seen in
Figs. 11a,b and 12. In the unfrustrated case 
$\alpha=0.0$ (open symbols), the $M$-distributions are smooth. Switching on the
frustration parameter $\alpha$ (solid symbols), we observe the emergence of a peak 
at $M=\frac{1}{6}$ in $S_{3}(\vec{p}=\pi (\frac{2}{3},\frac{2}{3}),M,\alpha)$ 
(Fig. 11a) and 
at $M=\frac{1}{3}$ in $S_{3}(\vec{p}=\pi (\frac{1}{6},\frac{5}{6}),M,\alpha)$ 
(Fig. 11b), respectively. 
The $M$-distribution of the transverse structure factor at $\vec{p}$=$(\frac{2}{3},
\frac{2}{3})\pi$ is shown in Fig. 12. The data for $\alpha=0.5$ show a break at
$M$=$0.42$, which reminds us to the structure observed in the one dimensional case
shown in Fig. 10b. 
%
\section{discussion and conclusion}
%
In this note, we have studied the zero temperature properties of the 
two dimensional spin-$\frac{1}{2}$ antiferromagnetic Heisenberg model in the presence
of an external field. Our results for the unfrustrated and the
frustrated model can be summarized as follows. \\
\\
1. The unfrustrated model : Owing to small finite-size effects and a weak dependence
on the boundary conditions, the magnetization
curve and the momentum  and field-dependence of the static structure factors can 
be determined with fairly good accuracy from the rather small systems (1.5),(1.6).
This statement holds away from the ''critical values'' $\vec{p}$=$(\pi,\pi)$, $M$=0.
At the momentum $\vec{p}$=$(\pi,\pi)$ the longitudinal structure factor is
finite for $M>0$, whereas the transverse structure factor is divergent if the
system size $N$ goes to infinite. At fixed magnetization $M$, the momentum dependence
of the static structure factor is smooth and well described by one variable, namely
(4.1). There is no indication for the existence of 
singularities, which might give a hint to field-dependent soft-modes - i.e. zero
frequency excitations. \\
\\
2. Frustration leads to new phenomena in the magnetization curve and
the static structure factors. The point $\alpha$=$0.5$ appears to be
of special interest. Here, the magnetization curve develops a square root
singularity (2.8) near saturation ($M\to \frac{1}{2}$). For smaller
values of $M$ - at $M$=$M_{0}(N)$ 
we find a shoulder in the magnetization curve and a dip in the
field-dependence of the transverse structure factor at $\vec{p}$=$(\pi,\pi)$.
However, the position $M_{0}(N)$ - where these phenomena appear - varies with
the system size and the boundary conditions. We do not know, whether these 
phenomena are finite-size effects or will survive in the 
thermodynamical limit.\\
\\
We looked for singularities in the field-dependence of the static
structure factors at fixed momenta. Indeed maxima emerge for momenta close
to $\vec{p}$=$(\pi,\pi)$ and $\vec{p}$=$(0,\pi)$,$\vec{p}$=$(\pi,0)$ if
the
system is frustrated sufficiently. They might originate from a soft-mode
with field-dependent momentum $\vec{p}(M)$ connecting 
$\vec{p}$=$(\pi,\pi)$ and $\vec{p}$=$(0,\pi)$, $\vec{p}$=$(\pi,0)$. 
However, our systems (1.5) (1.6) are much too small, in order to determine the
field-dependence of the trajectory  $\vec{p}(M)$. \\
\\
We have restricted our
considerations to a limited range $0\leq \alpha \leq \frac{1}{2}$ in the
frustration parameter $\alpha$. Going beyond $\alpha$=$\frac{1}{2}$ we
meet level crossings. E.g. the ground states in the sectors with $S_{z}$=$\frac{N}{2}
-1$ (called '' one magnon states'') have different momenta for $\alpha<\frac{1}{2}$
and $\alpha>\frac{1}{2}$, respectively. Level crossings are ''felt'' by the Lanczos 
algorithm, which meets increasing problems to find the ''true'' 
ground state among two competing states. The problem can be solved, if these
two states differ in their quantum numbers. One of the two competing states is
filtered out, if the starting vector has the appropriate quantum numbers. Therefore,
an extension of our analysis to the strong frustration regime $\alpha>\frac{1}{2}$
demands for a careful analysis of the ground state quantum numbers in the
sectors with a given total spin. We expect, that level crossings for $\alpha>
\frac{1}{2}$ will generate interesting phenomena. E.g. they might be responsible
for the plateau in the magnetization curve predicted by S. Gluzman. \cite{Gluzman94}
%
\acknowledgements
%
We got the numerical data for the static structure factor on the
$6\times 6$ system - which enter in Fig. 9 - from H.J. Schulz and the
field-dependence of the one dimensional structure factors in Figs. 10a,b from
M. Schmidt. We thank H.J. Schulz and M. Schmidt for their support and
S. Gluzman for discussions on the possible existence of plateaus in 
the magnetization curve.
We are indebted to M. Karbach and G. M\"uller for a critical reading of 
the manuscript and many useful comments. 
%
%

%
%
\newpage
\begin{figure}\label{fig1}
   \caption{The square lattice $5\times 5$+1=26 with helical boundary condition}
\end{figure}
\begin{figure}\label{fig2}
   \caption{Scaling behavior of the difference ratio (2.3) in the strong field 
    limit for the unfrustrated ($\alpha$=0.0 open symbols) and the frustrated case
   ($\alpha$=0.5 solid symbols). The curves represent polynomial fits in the
   scaling variable (2.4)}
\end{figure}
\begin{figure}\label{fig3}
     \caption{The magnetization curves computed on finite systems with 
      $N$=24,26,36,48 and 50 with the method of Bonner and Fisher [10] 
      a) the unfrustrated case $\alpha$=0.0 
      b) the frustrated case with $\alpha$=0.5}
\end{figure}
\begin{figure}\label{fig4}
   \caption{The field-dependence of the longitudinal structure factor 
   $S_{3}(\vec{p}=(\pi,\pi),M,\alpha,N)$ for $\alpha$=0.0 (open symbols) and
   $\alpha$=0.5 (solid symbols).The curves are shown to guide the eye}
\end{figure}
\begin{figure}\label{fig5} 
   \caption{The field-dependence of the transverse structure factor 
    $S_{1}(\vec{p}=(\pi,\pi),M,\alpha,N)$ for a) $\alpha$=0.0, b) $\alpha$=0.5}
\end{figure}
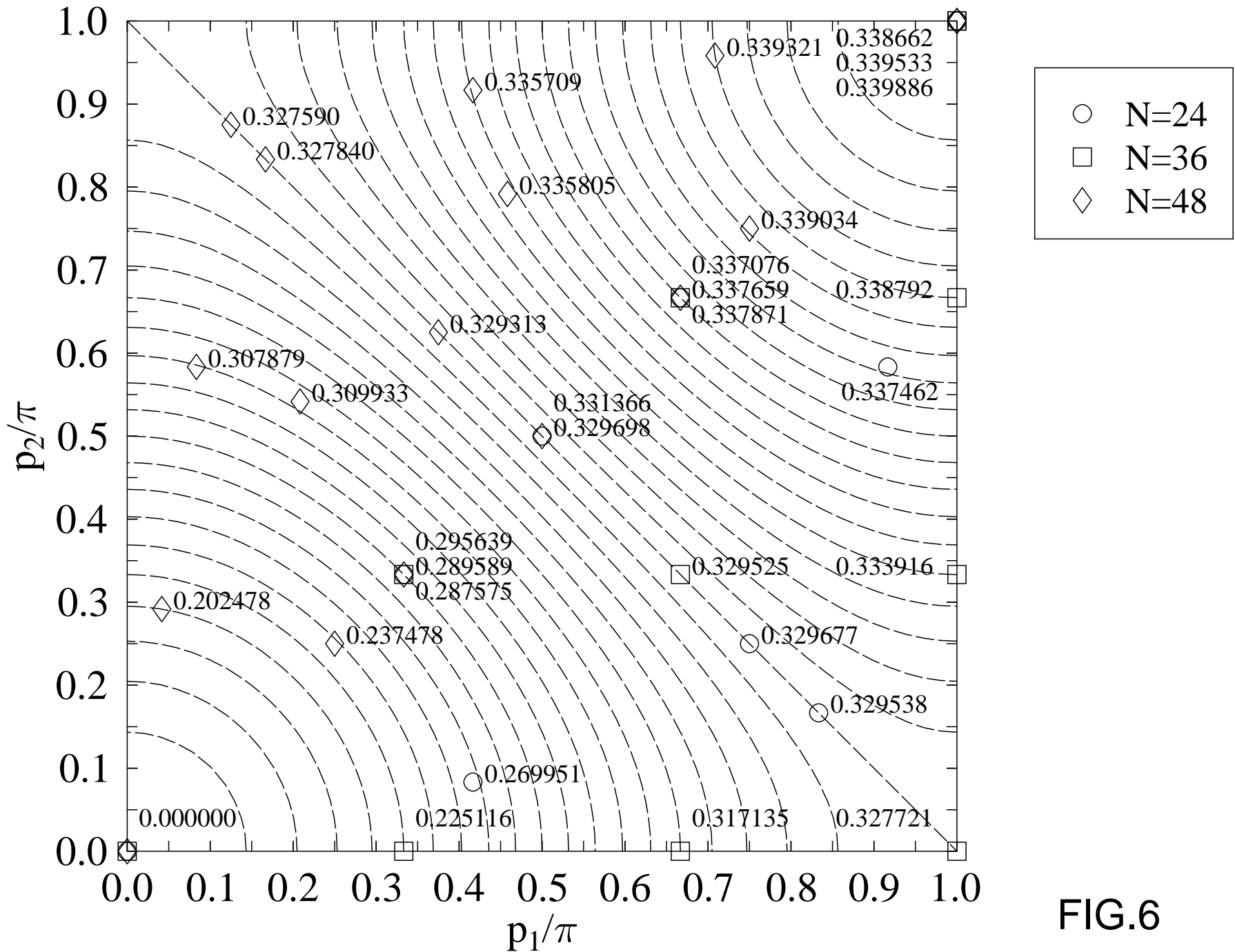
\begin{figure}\label{fig6}  
   \caption{The momentum dependence of the longitudinal structure factor  
    $S_{3}(\vec{p}=(p_{1},p_{2}),M=\frac{5}{12},\alpha=0.0,N)$ at fixed magnetization}
\end{figure}
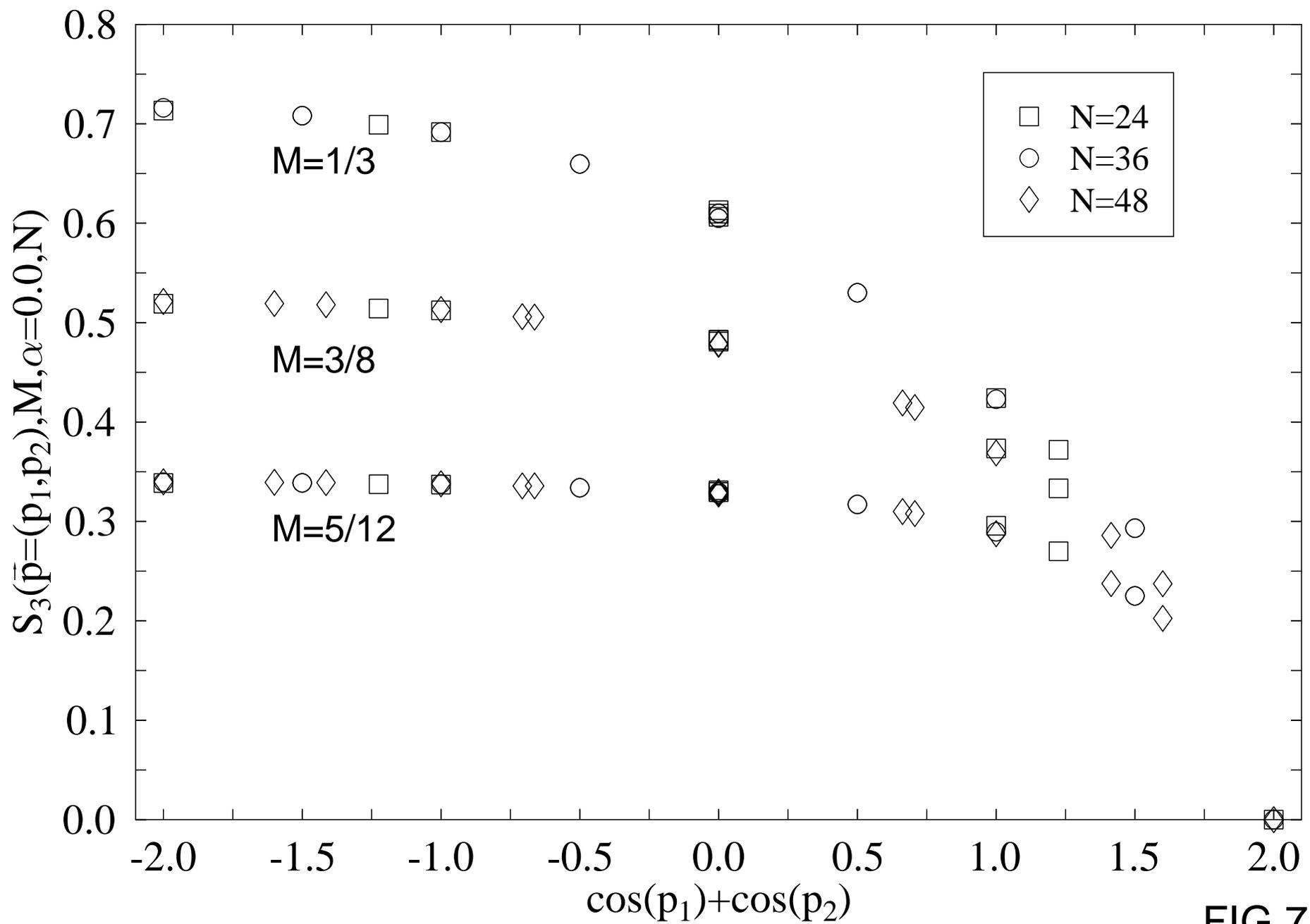
\begin{figure}\label{fig7} 
    \caption{Scaling of the longitudinal structure factor  
     $S_{3}(\vec{p}\!\!=(p_{1},p_{2}),M,\alpha=0.0,N)$ in the variable 
     $x$=$\cos{p_{1}}+\cos{p_{2}}$ at fixed $M$=$\frac{1}{3}$,$\frac{3}{8}$ and 
     $M$=$\frac{5}{12}$}
\end{figure}
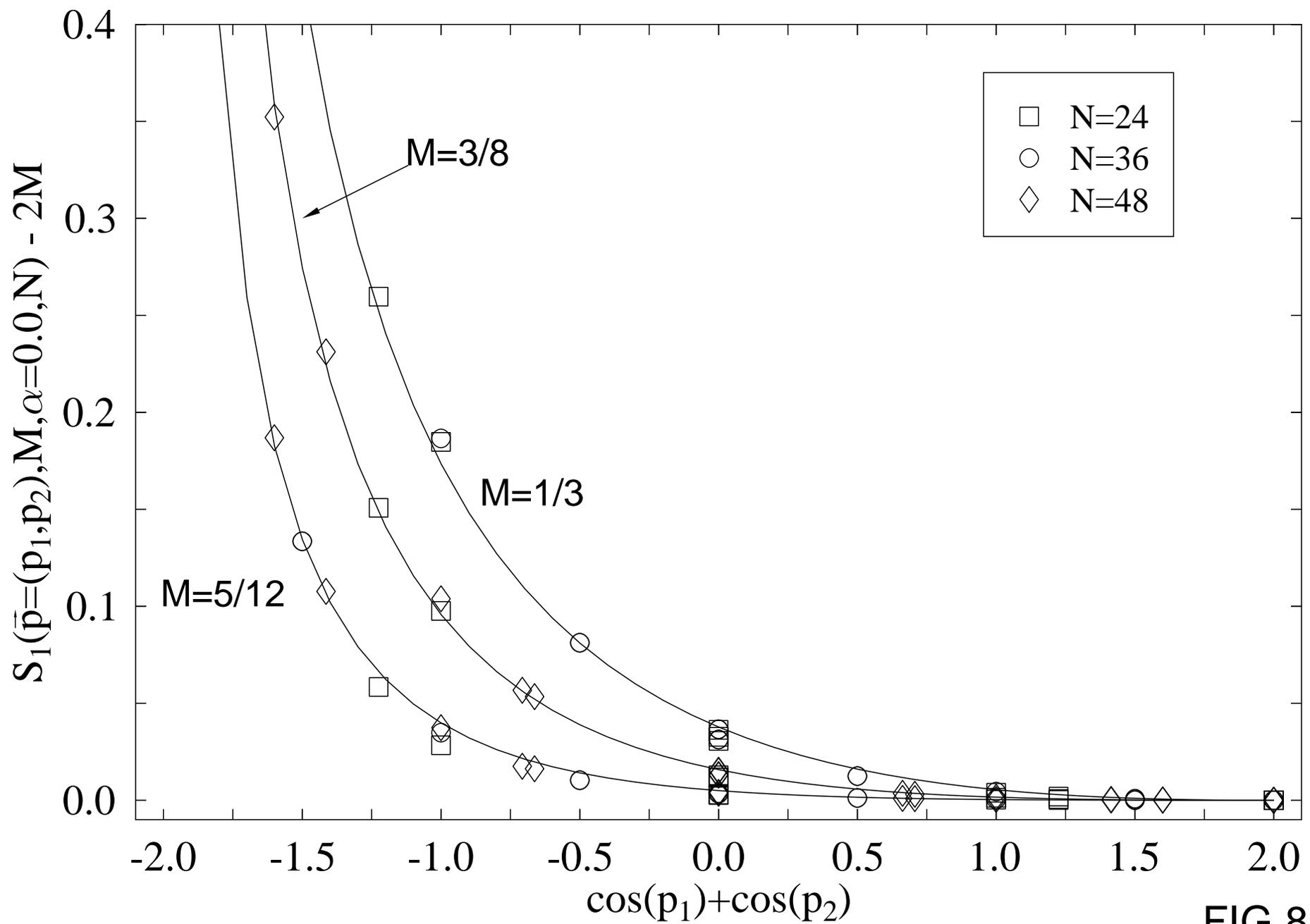
\begin{figure}\label{fig8} 
   \caption{Same as Fig. 7 for the transverse structure factor.
            The curves are shown to guide the eye}
\end{figure}
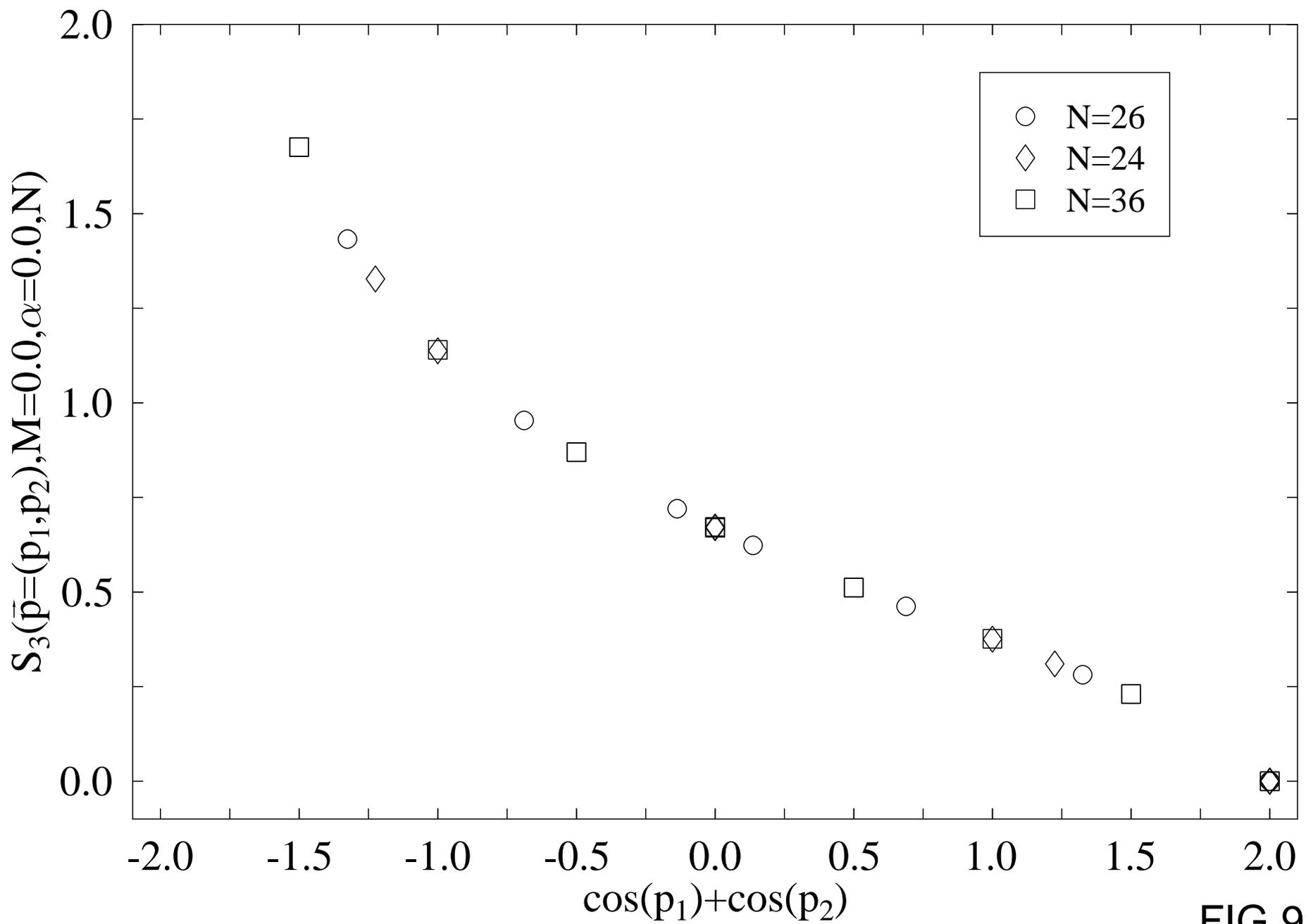
\begin{figure}\label{fig9}
   \caption{Same as Fig. 7 for M=0.0}
\end{figure}
\begin{figure}\label{fig10} 
    \caption{Field dependence of the one dimensional structure factors at fixed momenta
     $p$=$\frac{3}{4} \pi$, $\frac{2}{3} \pi$, $\frac{1}{2} \pi$, $\frac{1}{3} \pi$   
     a) the longitudinal case 
     b) the transverse case}
\end{figure}
\begin{figure}\label{fig11}  
    \caption{Field dependence of the two dimensional longitudinal 
     structure factors at fixed momenta
     a) $\vec{p}$=$\pi(\frac{2}{3},\frac{2}{3})$ 
     b) $\vec{p}$=$\pi(\frac{1}{6},\frac{5}{6})$  
open symbols $\alpha$=0.0 and solid symbols $\alpha$=0.5}
\end{figure}
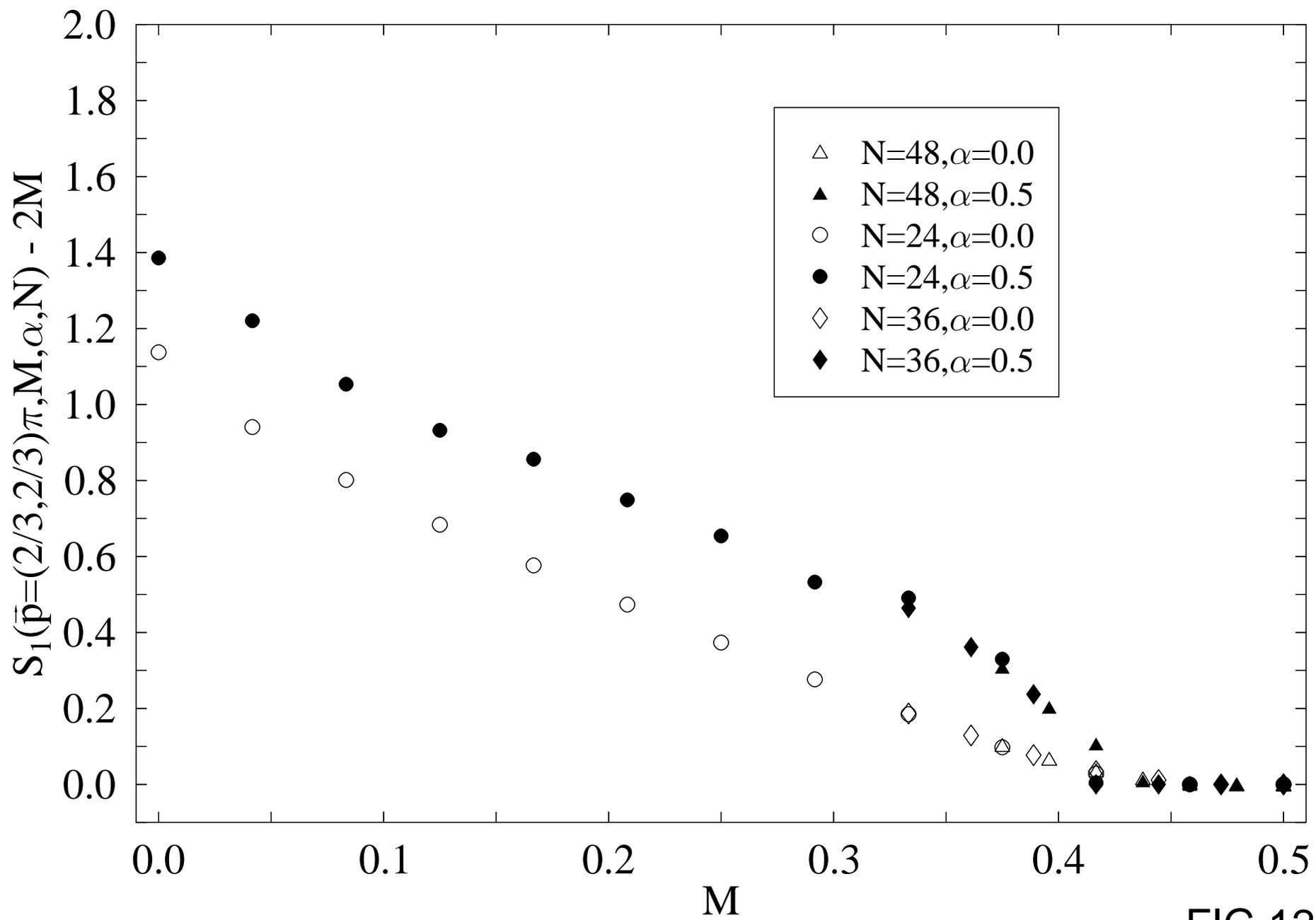
\begin{figure}\label{fig12} 
   \caption{Field dependence of the two dimensional transverse structure factor at fixed
    momentum $\vec{p}$=$\pi(\frac{2}{3},\frac{2}{3})$}
\end{figure}
%
%
\newpage
\begin{table}
\caption{The coefficients $D_{n}(\alpha)$ of the expansion (2.3)}
\begin{tabular}{c c c c c c c c c}   
$\alpha$               &  0    & 0.3   & 0.4   & 0.45  & 0.49  &  0.495 & 0.5 \\ \hline 
$D_{1}(\alpha)$ & 1.653 & 1.167 & 0.802 & 0.478 & 0.063 &-0.003 & -0.074 \\ \hline
$D_{2}(\alpha)$ & 1.374 & 1.086 & 1.162 & 1.763 & 3.115 &  3.376 & 3.663 \\ \hline
$D_{3}(\alpha)$ & 0.918 & 2.928 & 3.077 & 2.317 &-0.012 &-0.482 &-1.003  \\ 
\end{tabular}
\end{table}


\begin{thebibliography}{10}

\bibitem{Bednorz86}
         {{J.G.} Bednorz} and {{K.A.} M\"uller},
         Z. Physik. B {\bf 64}, 188 (1986).

         {{C.W.} Chu} et al,
         Phys. Rev. Lett. {\bf 58}, 405 (1987).

\bibitem{Doniach88}
         {S. Doniach} et al,
         Europhys Lett. {\bf 6}, 663 (1988).

         {M. Inui}, {S. Doniach} and {M. Gabay},
         Phys. Rev. B {\bf 38}, 6631 (1988).

\bibitem{Schulz92}
         {{H.J.} Schulz} and {{T.A.L.} Ziman},
         Europhys Lett. {\bf 18}, 355 (1992).

         {{H.J.} Schulz}, {{T.A.L.} Ziman} and {D. Poilblanc},
         J. Physique. {\bf 6}, 675 (1996).

         {{U.J.} Wiese} and {{H.P.} Ying}, 
         Z.Phys.B {\bf 93}, 147 (1994).

\bibitem{Johnson86}
         {{M.D.} Johnson} and {M. Fowler},
         Phys. Rev. B {\bf B 34}, 1728 (1986).

         {{J.B.} Parkinson} and {{J.C.} Bonner},
         Phys. Rev. B {\bf B 32}, 4703 (1985).

\bibitem{Karbach95}
         {M. Karbach}, {{K.H.} M\"utter} and {M. Schmidt},
          J. Phys.: Condens. Matter. {\bf 7}, 2829 (1995).


\bibitem{Fledder96}
         {A. Fledderjohann}, {C. Gerhardt}, {{K.H.} M\"utter}, {A. Schmitt} and {M. Karbach},
         Phys. Rev. B {\bf 54}, 7168 (1996).

\bibitem{Mueller81}
         {G. M\"uller}, {H. Thomas}, {G. Beck} and {{J.C.} Bonner},
         Phys. Rev. B {\bf 24}, 1429 (1981).

\bibitem{Schmidt96}
         {M. Schmidt}, {C. Gerhardt}, {{K.H.} M\"utter} and {M. Karbach},
          J. Phys.: Condens. Matter. {\bf 8}, 553 (1996).

\bibitem{Haan92}
         {O. Haan}, {{J.U.} Kl\"attke} and {{K.H.} M\"utter},
         Phys. Rev. B {\bf 46}, 5723 (1992).

         {{D.D.} Betts}, {S. Masui}, {N. Vats} and {{G.E.} Stewart},
         Can. J. Phys. {\bf 74}, 54 (1996).

         {{D.D.} Betts} and {{G.E.} Stewart},
         ''Estimation of zero temperature properties of quantum spin systems
  on the simple cubic lattice via exact diagonalization on finite lattices'' - preprint.

\bibitem{Gluzman94}
         {S. Gluzman},
         Phys. Rev. B {\bf 50}, 6264 (1994).

\bibitem{Bonner64}
         {{J.C.} Bonner} and {{M.E.} Fisher},
         Phys. Rev. A {\bf 135}, 640 (1964).

\bibitem{Lozovik93}
        {{Yu.E.} Lozovik} and {{O.I.} Notych},
        Solid St. Commun. {\bf 85}, 873 (1993).

\end{thebibliography}
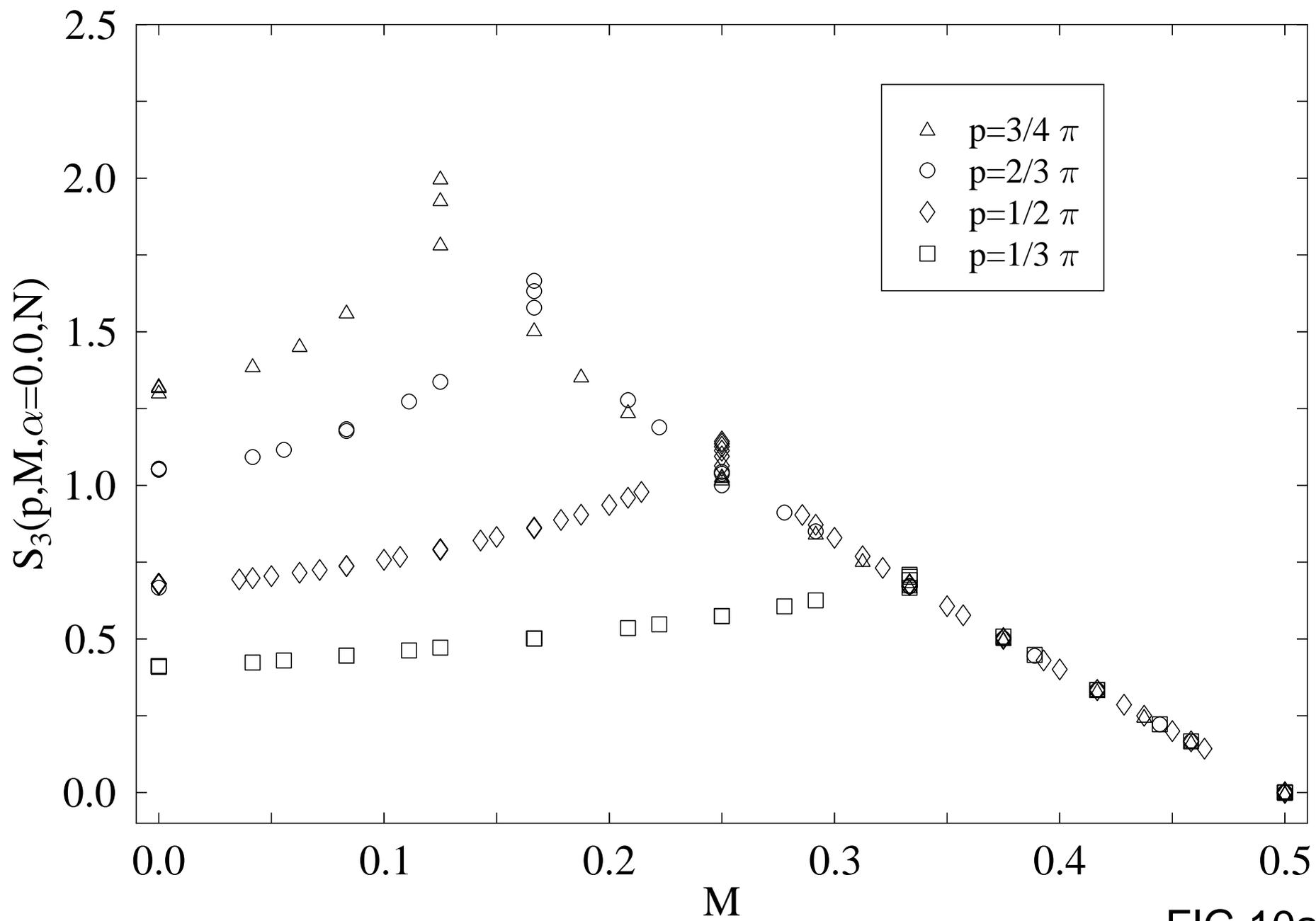
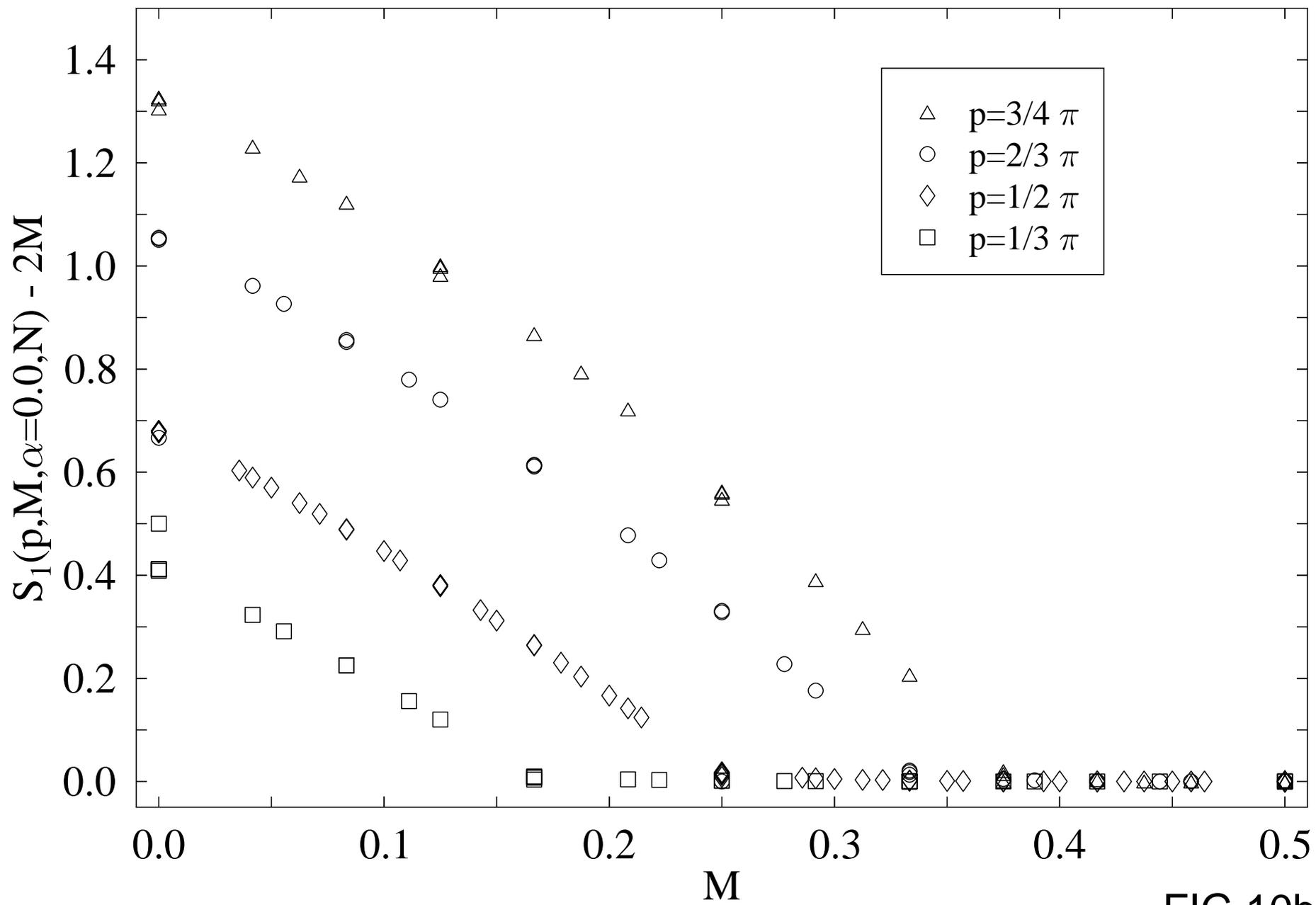
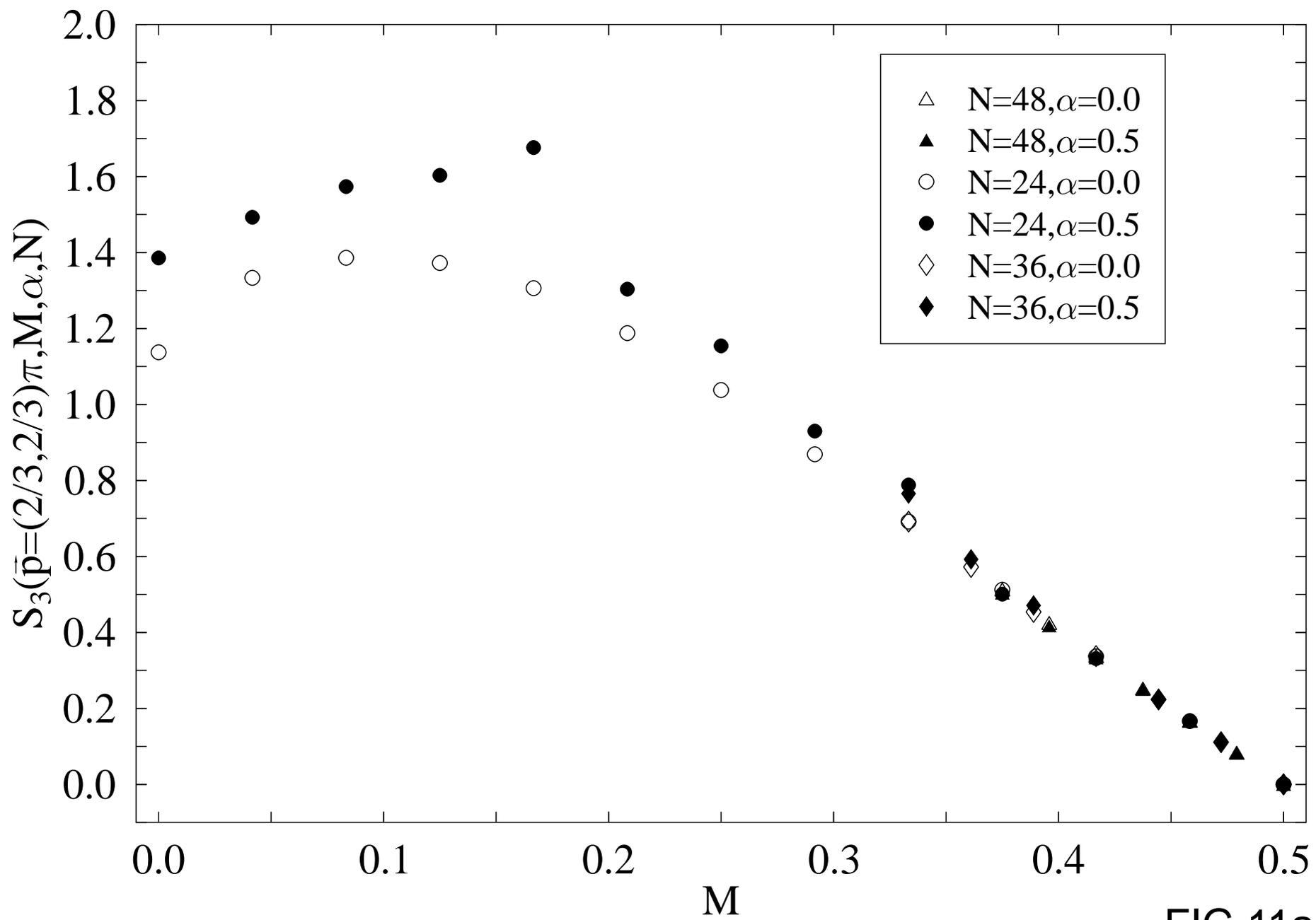
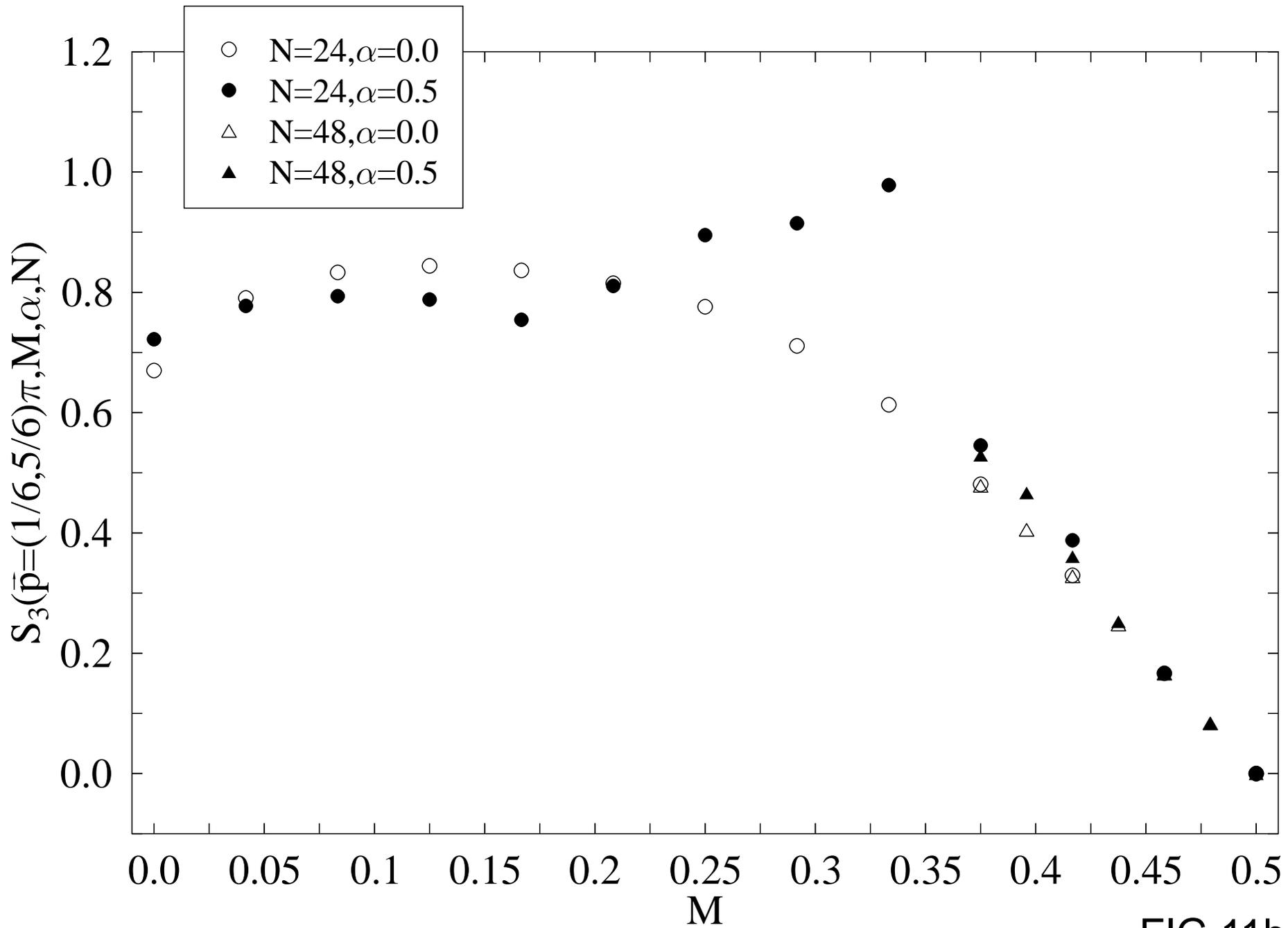
\end{document}